\documentclass[aps,prb,twocolumn,showkeys,floatfix]{revtex4-2}
\usepackage[hypertexnames=true]{hyperref}
\usepackage{textcomp,gensymb}
\usepackage{graphicx}
\usepackage{lipsum}
\usepackage[utf8]{inputenc}
\usepackage{newunicodechar}
\usepackage[utf8]{inputenc}
\DeclareUnicodeCharacter{0308}{\"{}}   % combining diaeresis → \" accent
\newunicodechar{⁺}{\textsuperscript{+}}

\begin{document}
	
\title{Optimized Broadband Cryogenic Ferromagnetic Resonance Spectrometer using a Closed Cycle Refrigerator
}

\author{Anna Merin Francis}
\author{Sunil Nair}
\affiliation{Department of Physics, Indian Institute of Science Education and Research, Pune, India}
%email{sunil@iiserpune.ac.in}
\date{\today}

\begin{abstract}

We present a vector network analyzer (VNA)-based broadband cryogenic ferromagnetic resonance (FMR) spectrometer, operating up to 20 GHz over a temperature range of 11–350 K. A cost-effective architecture is implemented through the integration of a closed-cycle refrigerator (CCR) and a custom-fabricated grounded coplanar waveguide (GCPW), designed for broadband transmission and reliable cryogenic operation. The VNA calibration is performed prior to measurements to account for microwave background and transmission losses, enabling reliable extraction of FMR spectra across the full temperature–frequency range. The sensitivity of the spectrometer is benchmarked using a yttrium-iron-garnet (YIG) thin film, yielding well-resolved resonances with narrow linewidths and high sensitivity.
\end{abstract}

%\keywords{}

\maketitle


\section{Introduction}
Ferromagnetic resonance (FMR) has long served as a technique for probing magnetization dynamics in ordered magnetic materials \cite{PhysRev.73.155,PhysRev.72.80}. By directly accessing the precessional response of the magnetization to a microwave excitation, FMR provides quantitative information on fundamental parameters such as the gyromagnetic ratio, effective magnetic anisotropy fields, and magnetic damping \cite{gurevich2020magnetization}. While FMR is a mature experimental tool, it has experienced a strong resurgence with the advent of spintronics \cite{pirro2021advances}, where the control and dissipation of spin angular momenta lies at the heart of device functionality \cite{barman20212021}. Within the framework of the Landau–Lifshitz–Gilbert (LLG) equation, FMR enables direct extraction of the Gilbert damping parameter, which governs the relaxation of magnetization dynamics and ultimately limits the efficiency of spintronic devices. In modern magnetic heterostructures, the damping is no longer an intrinsic material constant but acquires substantial contributions from interfacial spin-transfer processes, such as spin pumping and spin backflow    \cite{PhysRevB.81.214418,uchida2011long}, whose efficiencies are highly sensitive to temperature, electronic structure, and interfacial spin transparency  \cite{RevModPhys.25.239}. Under FMR excitation, the precessing magnetization acts as a dynamic source of spin current, injecting angular momentum into adjacent nonmagnetic layers, where it can be converted into a measurable transverse charge electromotive force via the inverse spin Hall effect (ISHE) \cite{PhysRevLett.111.217204}. Temperature-dependent FMR measurements are therefore essential not only for disentangling intrinsic and extrinsic damping mechanisms, but also for quantifying the efficiency of spin current generation and spin–charge conversion \cite{PhysRevLett.123.057203,PhysRevB.76.104409}, and for elucidating how spin orbit coupling, magnon–magnon scattering, and magnon-lattice interactions evolve across wide temperature ranges \cite{atsarkin2018temperature,heinrich1985fmr,PhysRevB.76.104416}.\\
In ultrathin films and multilayers, magnetic anisotropy often originates from competing bulk, surface, and interface contributions, each with it's distinct temperature dependence \cite{PhysRevB.80.180415}. Broadband FMR, when coupled with temperature control, provides a powerful approach for quantitatively disentangling these contributions, which are not accessible through static magnetometry alone \cite{artman1987magnetic}. This is particularly relevant for contemporary spintronic materials, where understanding the thermal evolution of dynamical parameters is essential for both fundamental insight and device optimization. Despite its importance, the experimental realization of broadband, temperature-dependent FMR studies remain technically challenging. Conventional cavity-based FMR lacks frequency agility \cite{willer2000s} and is poorly suited for quantitative damping analysis. Many broadband implementations are restricted to room temperature \cite{de2025openfmr} or rely on liquid-helium cryostats \cite{denysenkov2003broadband,montoya2014broadband}, which limit accessibility and operational flexibility. Although commercial broadband FMR systems capable of low-temperature operation are now available, their high cost limits widespread adoption. The CCR–based implementation provides a more practical alternative; however, the lowest achievable operating temperatures reported to date have been limited to approximately 27 K \cite{harward2011broadband} or even higher. In this work, we have optimized a broadband cryogenic FMR (CryoFMR) with wide temperature tunability and a cost-effective architecture for sensitive ferromagnetic resonance measurements. In addition, we implement a comprehensive VNA calibration procedure to correct the background contributions arising from various sources other than the device under test (DUT), a step that is absent in previously reported broadband FMR implementations. This calibration establishes a well-defined plane of reference at the sample, removing frequency-dependent losses and phase errors.
\vspace{-0.3cm}
\section{Instrumentation}
\begin{figure}
	\includegraphics[width=1\columnwidth]{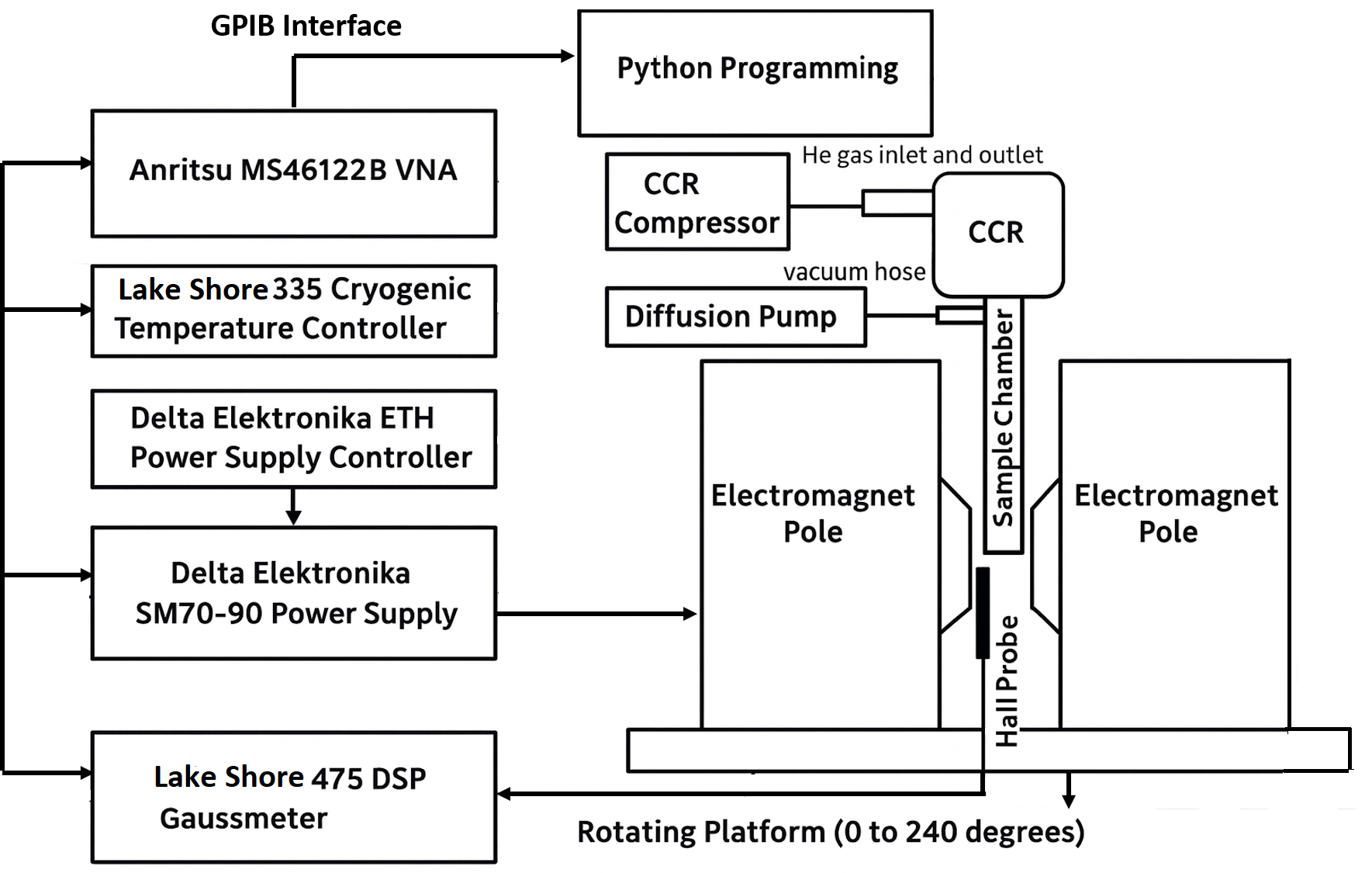}
	\caption{\footnotesize Block diagram of the Broadband CryoFMR spectrometer.}
	\label{CryoFMR}
\end{figure}
The schematic of the CryoFMR setup is shown in Fig.~\ref{CryoFMR}, comprising four main components: the vacuum system, DC magnetic field, temperature control, and microwave transmission. The vacuum system primarily comprises a diffusion pump, with the initial base pressure established using a rotary pump. A high-vacuum environment of the order of $10^{-6}$--$10^{-7}$ Torr is maintained in the sample space throughout the experiment to ensure efficient low-temperature operation. For the application of a DC magnetic field, an electromagnet mounted on a rotating platform is employed, powered by a high-voltage, high-current power supply capable of generating magnetic fields up to 0.6 T. The FMR measurements are performed in both frequency-sweep mode, with the magnetic field held constant, and field-sweep (continuous-wave) mode, with the excitation frequency fixed. The magnetic field can be rotated from in-plane to out-of-plane orientations, enabling FMR measurements in both geometries. The field sweeps are carried out with a step size of 0.1 Oe using a power controller integrated with the Delta Elektronika SM70-90 power supply, ensuring high field stability and precise control. This fine field resolution is particularly advantageous for measurements on samples exhibiting very low Gilbert damping, where the resonance linewidth is extremely narrow. The magnetic field is simultaneously monitored using a transverse Hall probe connected to a Lake Shore 475 DSP Gaussmeter, ensuring high sensitivity and precise field measurement with high decimal resolution. Since the transverse Hall probe is positioned closer to one of the magnet poles rather than at the sample location, a prior calibration was performed. In this procedure, two transverse Gaussmeter probes were used to measure the magnetic field simultaneously—one placed at the sample position and the other at the fixed probe location. From the linear relationship obtained between the two measurements, the magnetic field values recorded by the fixed probe were subsequently calibrated to yield the actual field at the sample position.\\
\begin{figure}
	\includegraphics[width=1\columnwidth]{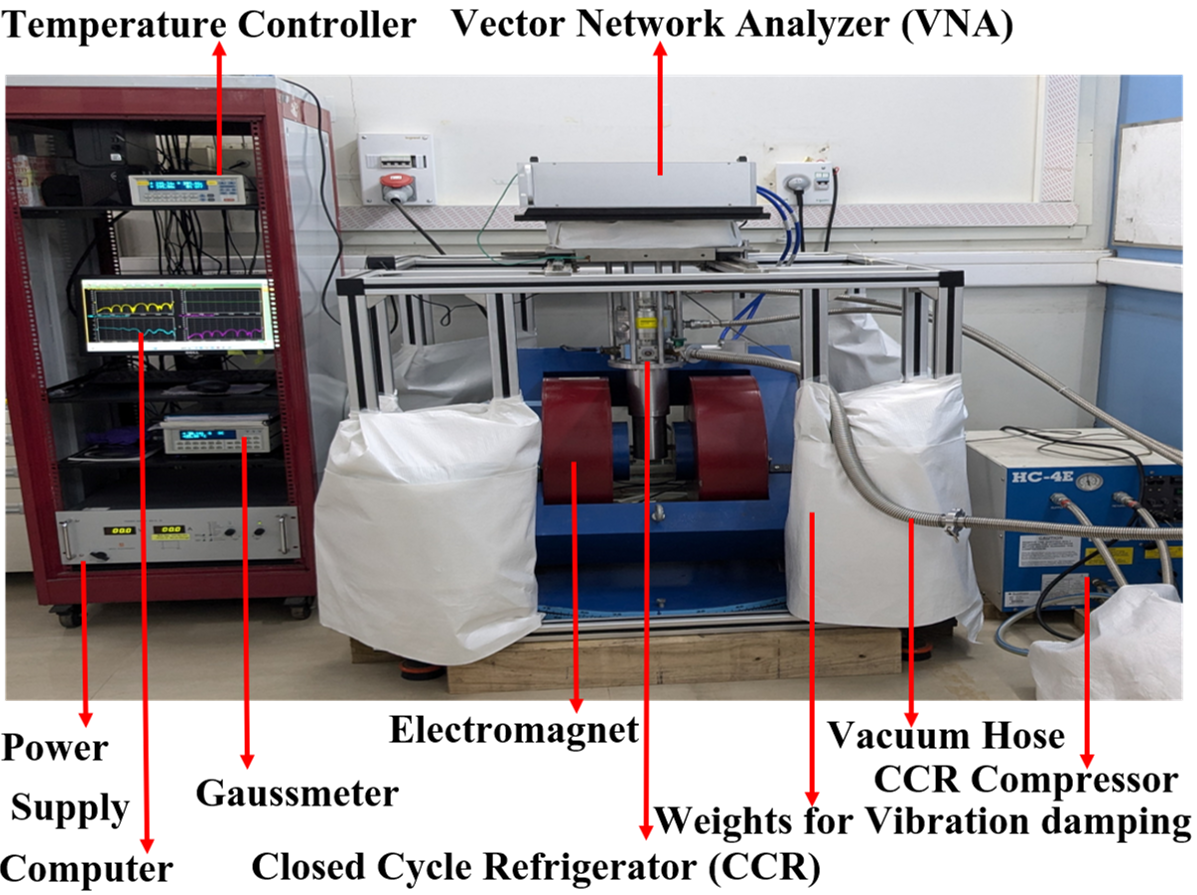}
	\caption{\footnotesize A picture of the in-house home-built broadband Cryo-FMR system with the CCR vacuum sample chamber placed between the poles of the electromagnet. The vacuum hose and compressor gas lines are kept along one direction to facilitate easy rotation of the electromagnet.}
	\label{Broadband CryoFMR}
\end{figure}
The temperature control unit comprises a Lake Shore 335 Cryogenic temperature controller equipped with two calibrated Cernox sensors—one mounted on the sample holder and the other on the waveguide, where the sample is mounted in a flip-chip configuration using Kapton tape, followed by black cloth tape for enhanced mechanical stability. A clear temperature difference ranging from 0.1 K to 1.5 K is observed between the readings of the two sensors. Hence, controlling the temperature using the Cernox sensor mounted on the waveguide provides a more accurate and reliable estimation of the actual sample temperature. In contrast, previously reported CryoFMR setups rely solely on temperature control via a sensor mounted on the sample holder. Incorporating the Cernox sensor on the waveguide, together with a cartridge heater positioned in close proximity to the waveguide, enables precise temperature regulation and allows for fine temperature step control when required.\\
The refrigeration system comprises a closed-cycle helium setup consisting of an HC-4E helium compressor and a CH-204 10 K cold-head refrigerator from Sumitomo Cryogenics. The refrigerator assembly, without the compressor, is mounted on an aluminum rack and carefully aligned to allow efficient electromagnet rotation while keeping the sample chamber between the poles. The rack is equipped with vibration-damping pads and additional mass at the corners to further reduce mechanical vibrations. The entire assembly is installed on a sliding platform, allowing it to be withdrawn from between the electromagnet poles for convenient sample loading and replacement. A layer of indium foil is inserted between the cold finger and the sample holder to enhance thermal anchoring and improve heat transfer. The removable vacuum shields are further modified according to custom specifications to ensure proper alignment of the sample chamber within the gap between the electromagnet poles. The image of the in-house home-built Broadband CryoFMR is shown in FIG.\ref {Broadband CryoFMR}.\\
\begin{figure}
	\includegraphics[width=1\columnwidth]{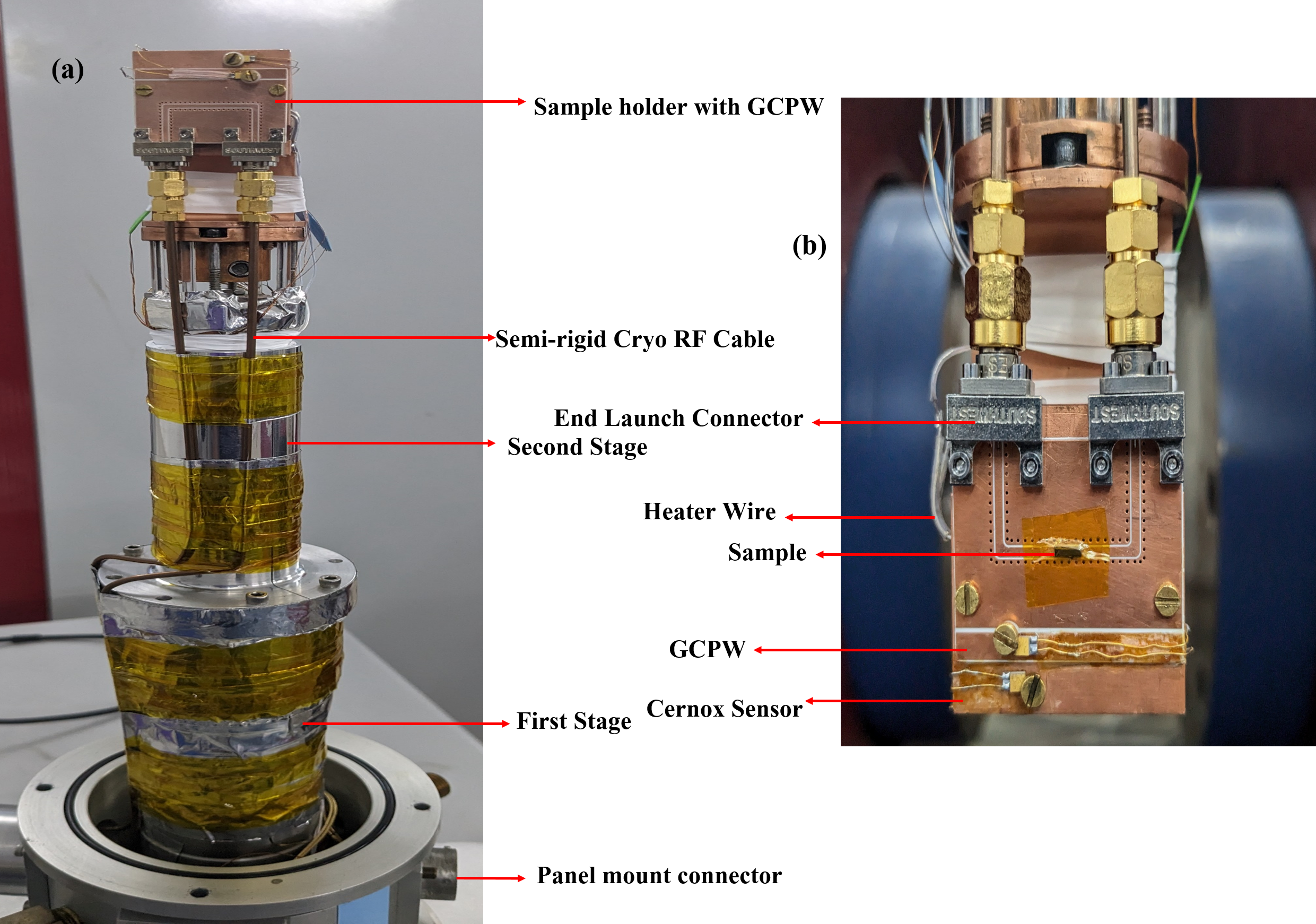}
	\caption{\footnotesize (a) A picture of the inverted CCR showing the cryogenic RF cable installation, with cable bending near the first- and second-stage junction, (b) Photograph of the copper sample holder with the grounded coplanar waveguide (GCPW) mounted on top, and the sample kept on the signal line in a flip-chip configuration using Kapton tape.}
	\label{SampleHolder}
\end{figure}
The microwave transmission unit is carefully selected and customized for low-temperature and high-frequency operation. The central component of the transmission line is a custom-fabricated, low-temperature-compatible GCPW \cite{simons2004coplanar}, designed to ensure reliable microwave performance under cryogenic conditions. A two-port vector network analyzer (the MS46122B Compact USB VNA) from Anritsu, with a maximum operating frequency of up to 20 GHz, is employed for both RF signal generation and detection. The microwave response is characterized through the measurement of scattering parameters, with primary emphasis on the transmission coefficient, expressed as transmittance in decibels (dB). Microwave transmission inside the vacuum chamber is achieved using non-magnetic semi-rigid Cryo RF cable assemblies supplied by CryoCoax-Intelliconnect, which are terminated with non-magnetic SMA connectors at both ends. The cables feature a silver-plated beryllium copper center conductor and a beryllium copper outer conductor (2.19 mm), providing mechanically robust, low-loss RF performance compatible with combined cryogenic and magnetic-field environments. To preserve impedance matching and minimize insertion losses, the cables are bent with strict adherence to the specified minimum bend radius using appropriate tooling. Controlled bends are introduced at two critical locations: (i) between the first and second stages of the CCR, and (ii) near the room-temperature (RT) hermetically sealed panel-mount SMA connector inside the vacuum chamber, which allows connection of the semi-rigid RF cable assembly to the external VNA RT SMA cable without compromising the vacuum. The semi-rigid RF cable assemblies are firmly attached to the CCR body using Kapton and aluminum tape in order to provide enhanced thermal anchoring, considering their relatively high thermal mass. This arrangement ensures good thermal stability and suppresses unwanted vibrations, thereby improving the signal-to-noise ratio. Alternatively, one can use flexible cryogenic RF cables—albeit at a higher cost—which eliminates the need for careful bending of RF cables during installation. An image of the sample holder, along with the wave guide and the bending of the RF cables inside the vacuum chamber, is shown in FIG. \ref{SampleHolder}.\\
\begin{figure}
	\includegraphics[width=1.05\columnwidth]{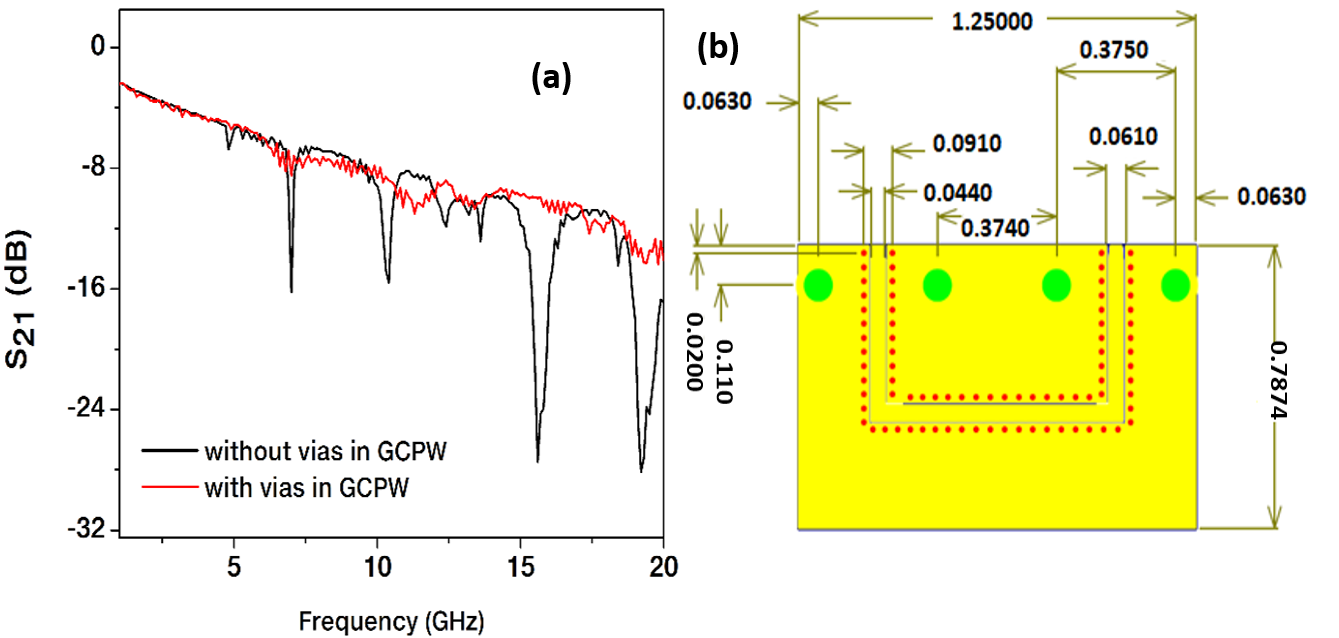}
	\caption{\footnotesize GCPW transmittance response as a function of frequency (a) with and without vias, (b)Top view of the GCPW showing dimensions in inches; the yellow color represents the copper cladding.}
	\label{vias}
\end{figure}
The RF signal is transmitted from the cables to the sample via a waveguide. For this purpose, we fabricated a GCPW utilizing a high-frequency, low-temperature-compatible dielectric material with copper cladding on both sides, specifically a 30-mil RO3003 laminate with 1 oz of copper, as provided by the Rogers Corporation. The RF cables are connected to the waveguide through non-magnetic end-launch connectors from Southwest Microwave, in which the connector core pin establishes pressure contact with the signal line rather than relying on soldering or silver paste bonding. The connector shield is mechanically fastened to the ground planes of the GCPW on both the top and bottom surfaces of the laminate.  The waveguide geometry, including the signal-line width and the gap between the signal and ground planes, was validated through impedance-matching calculations performed using Ansys High-Frequency Simulation Software (HFSS) to minimize reflections and maximize power transmission. The corresponding HFSS simulation data for the selected end-launch connector and laminate, spanning the required frequency range, was supplied by Southwest Microwave. The width of the signal line and the gap between the signal and the ground lines, obtained from HFSS, are 1117.6 $\mu$\text{m} and 432 $\mu$\text{m}, respectively, while the length and width of the GCPW are adjusted according to the sample holder and connector dimensions.\\
For copper cladding removal, an ammonia-based etching process was employed in accordance with standard PCB fabrication protocols, replacing conventional ferric chloride etching, to achieve improved dimensional fidelity and edge definition. Following the etching process, the GCPWs were systematically cleaned using diluted sulfuric acid and subsequently subjected to ultrasonication in isopropyl alcohol to eliminate residues. We observed an improvement in power transmission upon introducing vias near the signal line on the ground planes, compared to a plain GCPW. The vias were fabricated using a precision drill bit, followed by the deposition of copper along the via walls. The incorporation of vias enhances GCPW performance by improving electromagnetic field confinement \cite{sain2015impact}, suppressing parasitic modes, and ensuring a well-defined and continuous ground reference. Additionally, a slight conical (tapered) profile of the signal line near the pin connector in the GCPW is deliberately introduced to facilitate a smooth RF transition between the connector and the planar transmission line. This tapered region serves as a controlled transition, improving impedance matching, suppressing reflections, and thereby enhancing the overall transmission performance of the GCPW. The dimension of the entire GCPW, along with the vias, their positions, and the corresponding transmission responses, is shown in Fig.~\ref{vias}. Interestingly, the low-cost GCPW fabricated on a 30-mil RO3003 laminate exhibited better low-temperature stability than the commercially available GCPW made using a 10-mil RO4350 laminate, which is typically employed in most CryoFMR systems.\\
\begin{figure}
	\includegraphics[width=1\columnwidth]{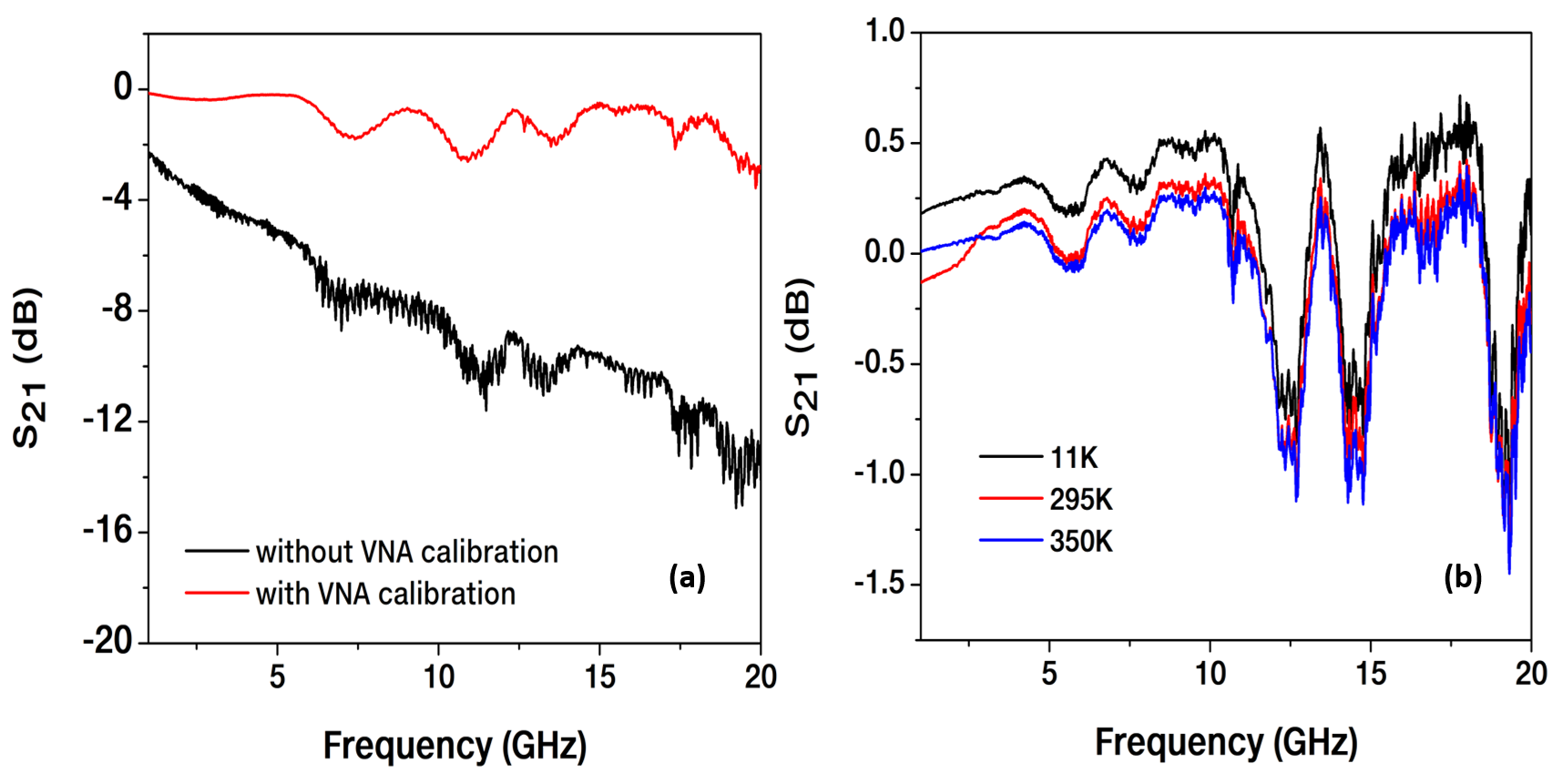}
	\caption{\footnotesize The GCPW response measured (a) with and without calibration, and (b) at different temperatures following SOLR calibration of the VNA performed at room temperature.}
	\label{VNA}
\end{figure}
A key aspect overlooked in earlier reports is that, unlike many other electronic instruments, the use of a VNA for RF signal generation and detection requires meticulous calibration prior to any measurement \cite{ReynosoHernandez2024VNA}. This calibration must be performed only after the RF assembly and the waveguide position have been firmly fixed, so that all systematic errors associated with the measurement setup are properly accounted for. The VNA calibration is fundamental to the precise extraction of the complex scattering parameters of a DUT, as uncalibrated measurements invariably contain systematic errors arising from RF cables, connectors, adapters, and the instrument's internal signal paths. Through calibration, these error contributions are accurately modeled and mathematically removed, effectively translating the measurement reference plane from the VNA ports to the actual terminals of the device. This procedure becomes especially crucial at high frequencies, where losses, impedance discontinuities, phase delays, and susceptibility to mechanical effects—such as cable movement or bending—can severely compromise both the magnitude and phase fidelity of the measured S-parameters.\\
For two-port measurements, a complete calibration is necessary to correct error terms associated with both reflection and transmission paths. The Short–Open–Load–Thru (SOLT) technique is therefore widely employed for a 2-port VNA, owing to its high accuracy and the availability of well-defined commercial calibration kits. In SOLT calibration, Short, Open, and Load are known reflection standards representing zero, infinite, and matched impedance, respectively, while Thru is a direct low-loss connection between the two VNA ports. However, in experimental configurations, repeated connection of a Thru standard or bending of RF cables is impractical and the SOLT approach can introduce additional phase and impedance errors due to mechanical perturbations. In such cases, the Short–Open–Load–Reciprocal (SOLR) calibration method is preferred, wherein the Thru standard is replaced by a passive, linear, and reciprocal network—also referred to as an unknown Thru—satisfying (S$_{21}$ = S$_{12}$), which remains fixed during calibration. In the FMR cryogenic RF assembly with the GCPW, the use of a standard Thru was impractical due to its fixed position and the additional RF cable bending that would be required. Therefore, SOLR calibration was performed at room temperature using a standard TOSLKF50A-20 Calibration Kit from Anritsu for the Short–Open–Load standards, and GCPW as the unknown reciprocal Thru, satisfying the condition S$_{21}$ = S$_{12}$. The calibration was performed using the in-built mathematical algorithms provided in the shockLine software.  After calibration, all measurement parameters, such as source power and frequency step size, remain fixed;  any change in these parameters necessitates recalling a different calibration set. Once VNA calibration is performed, the measured scattering parameters are corrected for all systematic errors of the measurement setup, so the reported S-parameters correspond to the intrinsic response of the DUT rather than the effects of the cables, connectors, or adapters. After VNA calibration is complete, a clear reduction in the scattering parameters is observed, as shown in Fig.~\ref{VNA}. Also a slight offset in GCPW response with temperature is noticed, arising from temperature-dependent variations in the cables and connectors. However, room-temperature background subtraction removes the dominant contribution, thereby allowing the detection of weak signals. Any subsequent change in the RF cable assembly or waveguide alignment would invalidate the calibration, necessitating its repetition to maintain measurement accuracy. The VNA operation is interfaced and controlled using the built-in ShockLine software designed specifically for the instrument. To minimize AC interference, the VNA is housed inside a Faraday cage and mounted separately on top of the CCR rack with vibration-damping pads.\\
Both measurements and subsequent data fitting were automated using Python programming, with all instruments interfaced through GPIB. During frequency-sweep measurements, background subtraction was performed by measuring S$_{21}$ at a magnetic field significantly different (either higher or lower) from the ongoing resonance magnetic field. To extract the full width at half maximum (FWHM) and the resonance field from the absorption spectra, the data were fitted using a Lorentzian function incorporating both symmetric and antisymmetric components \cite{petrakis1967spectral}. The symmetric Lorentzian component corresponds to pure microwave absorption and is directly associated with magnetic damping. In contrast, the antisymmetric Lorentzian originates from dispersive contributions to the measured signal, which can arise from phase differences between the RF magnetic field and the detected response, mixing of absorption and dispersion within the microwave detection chain, and imperfections in impedance matching or the presence of standing waves in the transmission line. The fitting model also includes a linear background term to account for a slowly varying non-resonant contribution to the spectra, arising from instrumental response and residual baseline drifts, thereby ensuring accurate extraction of the intrinsic FMR parameters. The absorption spectra were fitted using the following expression:
{\small
\begin{equation}
y(x) = y_0 + kx 
+ A \frac{\left(\frac{w}{2}\right)^2}{\left(\frac{w}{2}\right)^2 + (x - x_c)^2}
+ B \frac{\left(\frac{w}{2}\right)(x - x_c)}{\left(\frac{w}{2}\right)^2 + (x - x_c)^2}
\end{equation}
}
where $x_c$ denotes the resonance field, $w$ represents the linewidth (FWHM), A and B are the amplitudes of the symmetric and antisymmetric Lorentzian components, respectively, and $y_0 + kx$ describes the linear background contribution.
 One important aspect to note is that the absolute magnitude of the FMR signal depends sensitively on the coupling between the sample and the GCPW, which is governed by the thickness and uniformity of the air gap between them. Therefore, maintaining good adhesion of the sample to the waveguide over the entire temperature range is critical. Moreover, since the scattering parameters are presented on a logarithmic scale, the signal-to-noise ratio is not explicitly reported. For spectra affected by noise, a modest level of signal averaging can also be applied to obtain a cleaner absorption profile; however, this step is optional.
\begin{figure}
	\includegraphics[width=1\columnwidth]{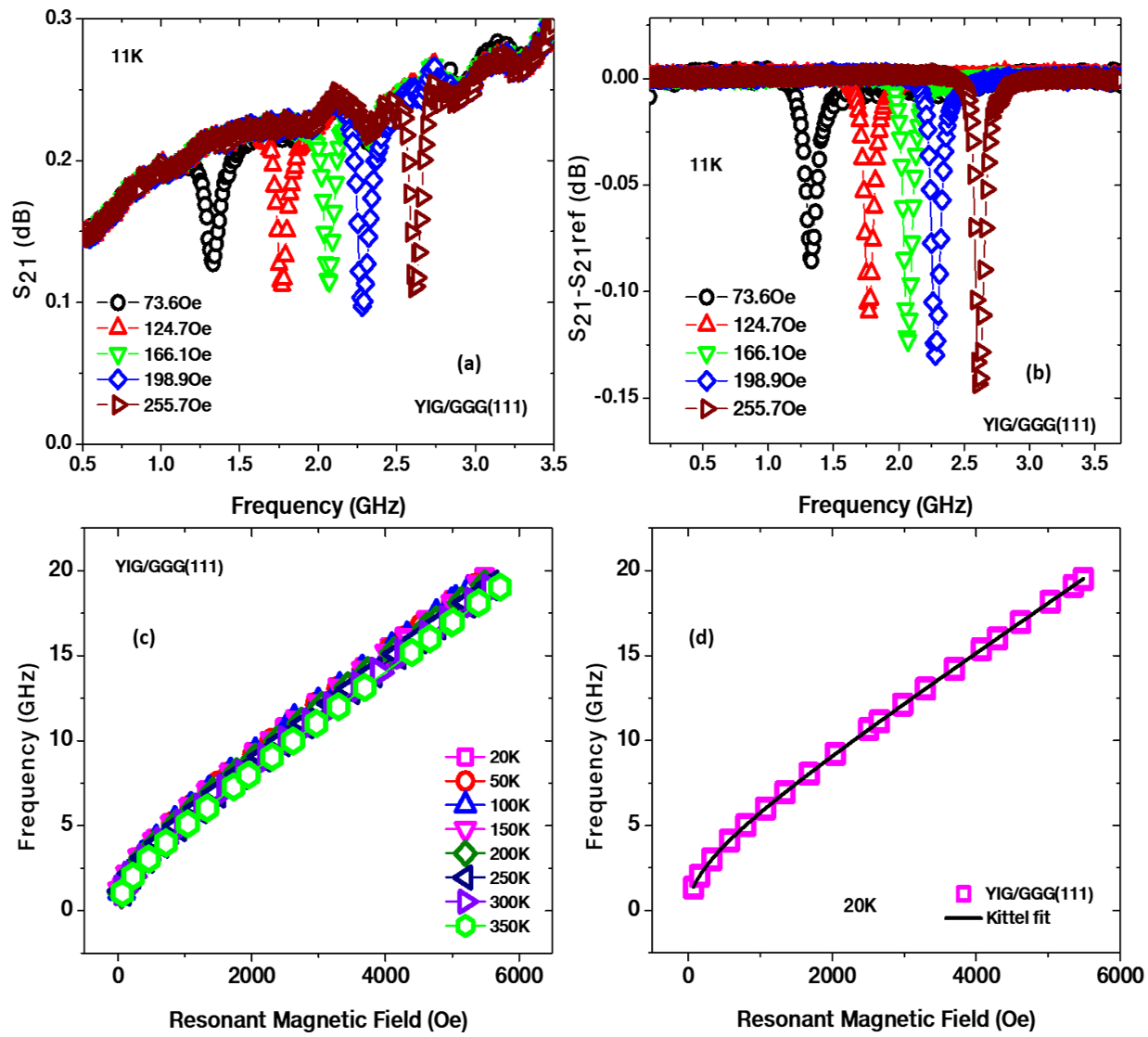}
	\caption{\footnotesize Frequency sweep measurements on YIG/GGG(111) thin film at 11 K (a) without background subtraction and, (b) with background subtraction using either a higher or lower resonance field as reference, (c) Frequency vs. Resonant Magnetic Field at various temperatures, (d) in-plane Kittel's fit }
	\label{frequencysweep}
\end{figure}
\section{Results and Discussion}
To establish the functionality and reliability of the home-built cryoFMR spectrometer, we conducted FMR measurements on a 100-nm-thick YIG thin film grown on a GGG (111) substrate by pulsed laser deposition. YIG is a benchmark material for magnonics \cite{yang2018fmr,sheng2021magnonics}, widely recognized for its exceptionally low Gilbert damping and long spin-wave propagation length. Frequency-sweep and field-sweep measurements were performed in the in-plane FMR geometry over a broad temperature range of 11–350 K and across frequencies spanning 1–20 GHz. The frequency-sweep measurements on the YIG thin film at 11 K, both with and without background subtraction, along with field-sweep measurements at 11 K and at various temperatures, are shown in Fig.~\ref{frequencysweep}. For each frequency and temperature, the field-sweep spectra were fitted using the Lorentzian equation as described earlier to extract the resonance field \(H_{\mathrm{res}}\) and the linewidth \(\Delta H\). The extracted resonance fields were subsequently analyzed using the in-plane Kittel formalism to determine the gyromagnetic ratio \(\gamma\) and the effective magnetization at each temperature. To extract the Gilbert damping parameter, the linewidth data were analyzed within the framework of the LLG equation. The frequency dependence of the linewidth is given by
\begin{equation}
\Delta H(f) = \Delta H_0 + \frac{4\pi \alpha f}{\gamma},
\label{eq:LLG}
\end{equation}
where \(\Delta H(f)\) is the frequency-dependent FMR linewidth, \(\Delta H_0\) represents the inhomogeneous broadening contribution, and \(\alpha\) is the Gilbert damping parameter. From the linear fit of \(\Delta H\) as a function of frequency \(f\), the slope yields \(\alpha\), while the intercept provides \(\Delta H_0\).
\begin{figure}
	\includegraphics[width=1\columnwidth]{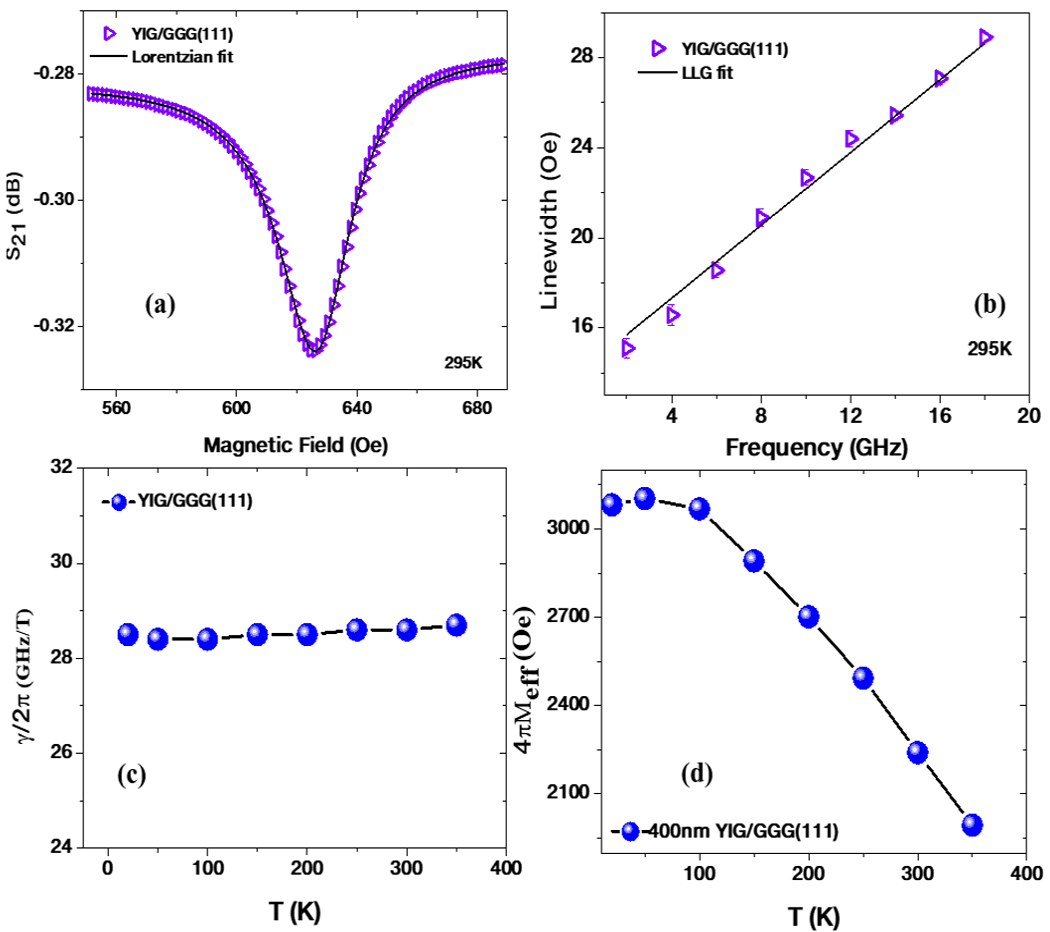}
	\caption{\footnotesize (a) Lorentzian fit and, (b) LLG fit of the YIG/GGG(111) thin film FMR data, (c) temperature-dependent gyromagnetic ratio and, (d) effective magnetization of the YIG/GGG(111) thin film. An increase in the effective magnetization is observed at the lower temperatures.}
	\label{parameters}
\end{figure}
The effective magnetization of the sample was determined by analyzing the frequency dependence of the resonance field using the in-plane Kittel equation,
\begin{equation}
f = \frac{\gamma}{2\pi}
\sqrt{H_{\mathrm{res}}\left(H_{\mathrm{res}} + 4\pi M_{\mathrm{eff}}\right)},
\label{eq:Kittel}
\end{equation}
where \(4\pi M_{\mathrm{eff}} = 4\pi M_s - H_{\mathrm{eff}}\), with \(M_s\) denoting the saturation magnetization and \(H_{\mathrm{eff}}\) the effective anisotropy field. By fitting the experimental data with Eq.~(\ref{eq:Kittel}), the gyromagnetic ratio \(\gamma\) and the effective magnetization \(M_{\mathrm{eff}}\) were extracted as functions of temperature. The temperature dependence of the measured gyromagnetic ratio and the effective magnetization are shown in Fig. \ref{parameters}. A constant gyromagnetic ratio value of 28.5 GHz/T confirmed it's alignment with the reported values \cite{haidar2015thickness}.  From the LLG fit, the Gilbert damping value of the YIG thin film at 295 K is $\alpha = (1.15 \pm 0.06)\times 10^{-3}$, confirming low Gilbert damping behavior \cite{PhysRevB.102.014411}. Our measurements also demonstrate the sensitivity of the instrument in picking FMR linewidths as low as 5 Oe, along with transmittance parameter achieved with a better resolution of 0.02 dB.\\
\section{CONCLUSIONS}
We have successfully built and calibrated a VNA-based broadband cryogenic FMR spectrometer operating over a temperature range of 11–350 K and frequencies up to 20 GHz. A detailed methodology for the custom fabrication and implementation of grounded coplanar waveguides is presented. The resulting system achieves the lowest reported operating temperature for a CCR-based cryogenic FMR platform. The implementation of systematic VNA calibration ensures accurate correction of microwave background and transmission losses, which is essential for reliable extraction of resonance fields, linewidths, and damping parameters over wide temperature and frequency ranges. The performance of the instrument is validated using a low-Gilbert-damping YIG thin film, where narrow linewidths and low Gilbert damping values confirm the high sensitivity of the setup. This cost-effective, versatile, and accessible platform enables quantitative investigations of temperature-dependent magnetization dynamics, interfacial spin-transfer efficiency, and magnetic anisotropy in thin films and heterostructures, providing a valuable experimental tool for fundamental studies and spintronic device research.

\begin{acknowledgments}
\ A. M. F. acknowledges Satish Lokhande and Pravin Raybole (GMRT, TIFR) for their help during the installation of the cryogenic RF cables. A.M.F. also thanks Ganesh Pawar for assistance with Python programming and Harshdip Meshram for his support. A.M.F. acknowledges DST-INSPIRE for providing financial support through a Senior Research Fellowship (SRF). A.M.F. and S.N. acknowledge funding support by the Ministry of Human Resource Development (MHRD) through the Scheme for Transformational and Advanced Research in Sciences (STARS).\\

\end{acknowledgments}
\bibliography{manuscript}

\end{document}